\documentstyle[11pt,aaspp4]{article}
\slugcomment{Accepted for publication in ApJ Letters}

\lefthead{Nakajima et al.}
\righthead{Methane in an L dwarf}

\begin{document}

\title{$H$ and $K$ Band Methane Features in an L Dwarf, 2MASS 0920+35\altaffilmark{1}}

\author{Tadashi Nakajima\altaffilmark{2}}
\affil{National Astronomical Observatory, 2-21-1 Osawa, Mitaka,
181-8588, Japan}

\author{Takashi Tsuji}
\affil{Institute of Astronomy, The University of Tokyo, 2-21-1 Osawa, Mitaka,
181-0015, Japan}

\and

\author{Kenshi Yanagisawa}
\affil{Okayama Astrophysical Observatory, Kamogata, Okayama, 719-0232, Japan}

% Notice that each of these authors has alternate affiliations, which
% are identified by the \altaffilmark after each name.  The actual alternate
% affiliation information is typeset in footnotes at the bottom of the
% first page, and the text itself is specified in \altaffiltext commands.
% There is a separate \altaffiltext for each alternate affiliation
% indicated above.

\altaffiltext{1}{Based on data collected at Subaru Telescope, 
which is operated by the National Astronomical Observatory of Japan}
\altaffiltext{2}{tadashi.nakajima@nao.ac.jp} 

% The abstract environment prints out the receipt and acceptance dates
% if they are relevant for the journal style.  For the aasms style, they
% will print out as horizontal rules for the editorial staff to type
% on, so long as the author does not include \received and \accepted
% commands.  This should not be done, since \received and \accepted dates
% are not known to the author.

\begin{abstract}
We have obtained near-infrared spectra of three L dwarfs discovered
by 2MASS,
1506+13 (L3), 1507-16 (L5), and 0920+35 (L6.5). From the comparison
of the $H$ and $K$ band spectra of these L dwarfs, we have found
the presence of methane absorption in 0920+35. 
This implies that detectable methane absorption in the H and K
bands, usually considered the signature of a T dwarf, can be
present in objects classified optically as late L.
Methane detection in L dwarfs is consistent with 
the presence of a dust layer deep in the atmosphere as the unified
model of Tsuji suggests.

\end{abstract}

% The different journals have different requirements for keywords.  The
% keywords.apj file, found on aas.org in the pubs/aastex-misc directory, 
% contains a list of keywords used with the ApJ and Letters.  These are 
% usually assigned by the editor, but authors may include them in their 
% manuscripts if they wish. 


% That's it for the front matter.  On to the main body of the paper.
% We'll only put in tutorial remarks at the beginning of each section
% so you can see entire sections together.

% In the first two sections, you should notice the use of the LaTeX \cite
% command to identify citations.  The citations are tied to the
% reference list via symbolic KEYs.  We have chosen the first three
% characters of the first author's name plus the last two numeral of the
% year of publication.  The corresponding reference has a \bibitem
% command in the reference list below.
%
% Please see the AASTeX manual for a more complete discussion on how to make
% \cite-\bibitem work for you.   

\section{Introduction}

% Tsuji's prediction

% Methane in Jupiter and Gliese 229B

% Methane in T dwarfs and L/T transition objects

% L band methane found by noll et al.

% Delfosse et al. found indication of methane

% Interpretation by Tokunaga and Kobayashi

% Influence of Tokunaga and Kobayashi --- Reid et al.

% New interpretation

Companion searches around nearby stars
discovered 
first two  ultracool dwarfs which 
belong to 
spectral types later than M. 
In 1988, a brown dwarf candidate, GD 165B, was found as a companion to
a white dwarf (\cite{bec88}), 
and finally in 1995, the first cool brown dwarf, Gl
229B,
was found as a companion to a young M dwarf (\cite{nak95}).
Starting from 1997,
the number of ultracool dwarfs was 
significantly increased by the discoveries by sky surveys,
DENIS (\cite{del97}), 
2MASS (\cite{kir99}, \cite{bur99}), 
and SDSS (\cite{str99}) and
now ultracool dwarfs form two distinct spectral types 'L' and 'T'
(\cite{kir99}). L dwarfs which are
cooler than M dwarfs are selected by red color in the visible
or near infrared ($J-K >$ 1.3) and their subclasses are
defined by optical red spectra. 
T dwarfs which are even cooler than
L dwarfs are selected by very red color in the visible or
very red visible to infrared color and
defined by the presence of methane in the near-infrared
spectra. T dwarfs are blue in the near infrared ($J-K$ $\approx 0$).
As of writing of this letter, more than one-hundred L dwarfs 
and more than twenty T dwarfs are known in the literature,
including L/T transition objects (\cite{kir00}, \cite{bur01}, \cite{leg00}).
The L/T transition objects or early T dwarfs show both
methane and carbon monoxide in their $K$ band spectra and
have $J-K$ colors intermediate between L and T dwarfs.

From the point of view of observations, the sequence 
from L to early T to T dwarfs  appears natural
and there are a certain number of sample objects in 
each subclass of L dwarfs, early T dwarfs, and later T dwarfs.
However, there are issues related
to the L/T transition that need to be resolved before we fully understand
these ultracool dwarfs.

First of all, we need a theoretical explanation of 
the L/T transition process in which the behavior of
colors and spectra must be understood. 
In this paper, we compare the observed L dwarf spectra with
the unified models by Tsuji (2001). 
The models  explain the behavior of the infrared
colors through the sequence and approximately reproduce 
the spectra of L and T dwarfs.

Another issue is the presence of methane in L
dwarfs.  In the discovery paper of the first
DENIS L dwarfs, Delfosse et al. (2000) reported an apparent detection
of methane
in DENIS 0205-11AB (L7V) in the $K$ band.  However, 
Tokunaga and Kobayashi (1999) 
obtained a spectrum of this object and concluded that the 2 $\mu$m feature
was due to H$_2$ collision induced absorption (H$_2$ CIA).
Since then no researchers have claimed the presence of
methane in late L dwarfs in the $K$ band (\cite{kir99},\cite{rei01}).
On the other hand, the presence of methane in L dwarfs was confirmed
by L band spectroscopy of a L5V and L8V by Noll et al. (2000).
So methane exists in L dwarfs and it is worth revisiting
the features in late L dwarfs in the $H$ and $K$ bands taking into
account the possible presence of methane absorption.

\section{Observations}

Observations were carried out at the Subaru telescope on 2001 March 3
and 4 UT using the grism mode of the Infrared 
Camera and Spectrograph (IRCS) (\cite{kob00}). 
The slit width of $0.\hspace{-2pt}''6$ was sampled at $0.\hspace{-2pt}''058$
per pixel and the resolution was about 330 at $J$, $H$, and $K$.
The targets were nodded along the slit and observations taken
in ABBA sequence where A and B stand for the first and second
positions  on the slit.

2MASSW J1146345+223053 (henceforth 2M1146) was observed on March 3 UT.
At each of the four nod positions,
120 s exposures were acquired and the ABBA sequence was
repeated three times at $J$, 
With 120 s exposures, the sequence was repeated twice at $H$
and one cycle of 300 s exposures was obtained at $K$.
Immediately after the observations of 2M1146, $JHK$ spectra of
SAO81983 (G5V) were obtained for the calibration of telluric
transmission.
  
2MASSW J0920122+351742 (2M0920) was observed on March 4 UT.
The ABBA sequence was repeated twice at $J$ and
$H$
with 120 s exposures and
one cycle of  300 s exposures was obtained at $K$.
Immediately after the observations of 2M0920, $JHK$ spectra of
SAO61451 (G0V) were obtained for transmission calibration. 

2MASSW J1507476-162738 (2M1507) was observed on March 4 UT.
The ABBA sequence was obtained once at $J$ and
$H$ with 120 s exposures,
and  at $K$ with 150 s exposures.
Before the observations of 2M1507, $JHK$ spectra of
SAO159054 (G2V) were obtained for transmission calibration.

Data reduction was carried out using IRAF. 
Argon lines were used for wavelength calibration.
Telluric absorption was removed by dividing each object spectrum
with the corresponding 
 G star spectrum after  stellar absorption features had been removed.
The result was multiplied by a blackbody spectrum 
whose temperature is that of the effective
temperature of the G star.
For flux calibration 2MASS magnitudes were used
to normalize the spectra. 
Boxcar smoothing of 11 pixel wide (slit width) was applied
before obtaining the final spectra.

IRCS is designed so that optical throughput of the wavelength
regions where atmospheric transmission is poor is low. 
Usable wavelength regions are 1.18-1.35 $\mu$m, 1.49-1.81 $\mu$m,
and 1.95-2.4 $\mu$m respectively at $J$, $H$, and $K$.
Final spectra are shown in Figure 1.

\placefigure{fig1}

\section{Detection of methane features}

\subsection{2M0920}

In Figures 2 and 3, $H$ and $K$ band spectra are normalized so that
the portions of the spectra which are least affected by 
molecular absorption features overlap. To emphasize the locations
of H$_2$O and CH$_4$ absorption features, the spectra of Gl 229B
(\cite{geb96}) are also shown. 

\placefigure{fig2}

\placefigure{fig3}

In Figure 2, flux depression of 2M0920 (solid line) is conspicuous compared
to the other two objects and the absorption maxima at 1.63 and
1.67 $\mu$m due to
CH$_4$ 2$\nu_2+\nu_3$ and $2\nu_3$ bands are seen.
These absorption features 
 are also found in Gl 229B and in the early T dwarf, SDSS 1254-0122
(\cite{leg00}). 
In Figure 3, flux depression of 2M0920
due to methane  extending from 2.15 $\mu$m to
longer wavelengths is obvious as is the 2.20 $\mu$m absorption
maximum of the $\nu_2+\nu_3$ band.

\subsection{DENIS 0205-11AB}

Since we have seen the methane features in an L6.5 dwarf, it is natural
to revisit spectra of other L dwarfs. DENIS 0205-11AB (L7V)
was the first L dwarf for which methane detection was suggested (\cite{del97}).
 DENIS 0205-11AB is the coolest of the first three L dwarfs discovered 
by DENIS. Tokunaga and Kobayashi (1999)
 obtained a $K$ band spectrum of this object,
confirmed the flux depression longward of 2.2 $\mu$m, but interpreted
the feature as due to H$_2$ CIA.
This object was also observed by Leggett et al. (2001)  
and Reid et al. (2001), but
none of the authors mention the methane detection. Instead
Reid et al. (2001)  comment on the influence of H$_2$ CIA on
late L dwarfs referring to Tokunaga and Kobayashi (1999).

Now we question the interpretation of the 2.2 $\mu$m feature based on
H$_2$ CIA. Wavelength dependence of absorption
 produced by H$_2$ CIA was calculated
by Borysow and Frommhold (1990).
 One of the characteristics of this absorption 
is that it is spectrally broad. The absorption feature is not apparent within
a narrow spectral coverage unlike the 2.2 $\mu$m feature. 
If present, it should produce a large flux ratio between for example $H$
and $K$ bands, but should not cause an abrupt change within the $K$ band.
Here we conclude that the interpretation of the 2.2 $\mu$m feature
as methane by Delfosse et al. (1997) was correct. 

\subsection{Other L dwarfs}

Reid et al. (2001)
present the $JHK$ spectra of 2MASS 0310+16 (L8V) (Figure 8
of
their paper). 
In the $K$ band spectrum of this object, the 2.3 $\mu$m CO bandhead
is very weak while the 2.2 $\mu$m feature is strong. The weakness of
CO indicates that much of carbon is in CH$_4$. Reid et al.
 interpret
the 2.2 $\mu$m feature as due to H$_2$ CIA, but we argue that the
feature is naturally explained by methane.  

Kirkpatrick et al. (1999)
 present the $K$ band spectra of 2MASS 0850+10 (L6V) and
2MASS 1632+19 (L8V) (Figure 5 of their paper). The 2.2 $\mu$m
feature is apparent especially in 2MASS 0850+10, but they deny
the presence of methane. We again argue that the natural
interpretation
of this feature is methane absorption.

\section{Discussion}

\subsection{Complication in classification}

The detection of methane in the $H$ and $K$ bands of 2M0920 complicates
the definition of
the L/T transition. 2M0920 was classified as L6.5V by Kirkpatrick et
al. (2000)
based on red optical spectroscopy. 
On the other hand, Leggett et al. (2000)
 suggested that L/T transition objects
which show methane in the $H$ and $K$ bands should be classified as early T
dwarfs.
The difficulty may be arising from the fact that while L dwarfs are
defined in the optical wavelengths, T dwarfs are defined in the near
infrared. 

\subsection{Formation of methane bands in L dwarfs}

Methane can be formed at low temperatures
near 1000K under high densities, and detection of methane in
Gl 229B was deemed as evidence for very low  $T_{\rm eff} $ of this
object (\cite{opp95}).   
It was somewhat unexpected that methane was detected in
L dwarfs by Noll et al. (2000), even if the methane 
bands detected was the strong $\nu_{3}$ fundamental. It may appear to be
more surprising that the weaker combination bands of methane 
at 1.6 and 2.2 $\mu$m are now detected in an L dwarf as shown in \S3.
Such new observations, however, may imply that our understanding of
the photospheric structure of L dwarfs should radically be
reconsidered.

It was generally thought that the photospheres of L dwarfs are
dusty. However, simple dusty models of case B (\cite{tsu00}),
in which dust grains exist throughout
the photosphere so long as the thermodynamical condition of
condensation is fulfilled, cannot explain the formation of
the methane bands at all as shown in Fig.4a. In fact, no trace of
methane bands appears in any models of $T_{\rm eff} $ between 1500 and 
1800K. The reason for this is
that the strong extinction by dust masks the molecular bands on one hand
and also the backwarming effect of dust grains makes the photosphere
too warm for methane to form. On the other hand, 
dust segregated models (case C), in which dust once formed has precipitated
below the photosphere and volatile molecules dominate the observable
photosphere, easily show methane bands at 2.2 $\mu$m  as shown in Fig.4b.
In this case, weak methane absorption appears already at $T_{\rm eff} = 
1800$K and turns to be quite strong at the lower $T_{\rm eff}$'s 
(note
that the solid and dotted lines in Fig.4 represent the cases with and
without methane opacity, respectively). 
However, it is known that the infrared colors of such models are rather 
blue and incompatible with the observed red colors of L dwarfs.
It is clear that none of these models of cases B and C
offers an explanation
for the presence of methane bands in L dwarfs.

\placefigure{fig4}

A new model was developed based on the idea that
warm dust should exist deep in the photosphere (\cite{tsu99}), and
the observed infrared colors of L dwarfs can reasonably be reproduced
by the unified models with a thin dust layer deep in the photosphere
(Tsuji 2001). In these models, both dust formation and segregation
processes are taken into account and the thin dust layer is generated
naturally.
The unified models are successful
because the thin dust cloud is situated in the observable
photosphere of L dwarfs and this fact explains why the infrared
colors of L dwarfs are so red compared with those of T dwarfs in which
the dust layer is situated too deep in the photosphere to give
observable effects.  
At the same time, volatile molecules can reside in the upper
photosphere above the dust cloud in L dwarfs and this opens a possibility
for methane bands to be formed.
We confirm that the methane bands near 2.2 $\mu$m  appear 
and can be predicted to be reasonably strong for models of 
$T_{\rm eff}$ below about 1600K as shown in Fig.4c.
Thus only the unified model with $T_{\rm eff}$ between 1500 and 1600K
offers a reasonable account for the
presence of methane and, at the same time,  for the red colors due to 
dust extinction.

\section{Concluding Remark}

Methane absorption features are present in the $H$ and $K$ band
spectra of 2M0920 (L6.5). The $K$ band methane feature is also seen
in some other L dwarfs found in the literature. 
We have shown that objects that are optically classified as
L can have methane absorption
at 1$-$2.5 $\mu$m.
The presence of
methane in L dwarfs is consistent with the prediction of
the unified models of
Tsuji (2001) in which the presence of a thin dust layer deep in
the atmosphere is considered. 

\acknowledgments

We thank H. Terada and
the staff of the Subaru Observatory for excellent support 
and the anonymous referee for useful comments.
This work is supported in part by Grant-in-Aid for Scientific Research
(no. 11640227).

\clearpage

\clearpage

\figcaption[fig1.eps]{JHK spectra. They are linearly transformed
by y = C$_1$ * F$_\lambda$ + C$_2$ to make the comparison easy.  \label{fig1}}

\figcaption[fig2.eps]{Comparison of $H$ band spectra. 
Three spectra are normalized at 1.58 $\mu$m where
molecular absorption has minimal effect. 2M0920 (solid line)
shows flux depression compared to 2M1507 (dash-dot) and
2M1146 (dotted). The spectrum of Gl 229B (\cite{geb96})
is also shown for reference. The 
absorption maxima of CH$_4$
$2\nu_2+\nu_3$  and $2\nu_3$ bands respectively at 
1.63 and 1.67 $\mu$m 
are indicated.  \label{fig2}}

\figcaption[fig3.eps]{Comparison of $K$ band spectra. 
Three spectra are normalized at 2.08 $\mu$m where
molecular absorption has minimal effect. 2M0920 (solid line)
shows flux depression compared to 2M1507 (dash-dot) and
2M1146 (dotted). The spectrum of Gl 229B (dotted, \cite{geb96})
is also shown for reference. The absorption maximum of 
CH$_4$ $\nu_2+\nu_3$ band at 2.2 $\mu$m 
is indicated.  \label{fig3}}

\figcaption[fig4.eps]{
a) Synthetic spectra ($F_{\lambda}$ in unit of 10$^{-14}$
erg cm$^{-2}$ sec$^{-1}$ cm$^{-1}$) based on dusty models in which
small dust grains form in LTE. 
The results with and without methane opacity show no difference,
that is no methane band can be predicted by these dusty models
for $T_{\rm eff}$ between 1500 and 1800K.
b) The same as Fig.4a but based on the dust segregated models in which
dust formed but precipitated below the photosphere.
The solid and dotted lines represent the cases with and without
the methane opacity. c) The same as Fig.4b but based on the unified
models in which dust forms only  in a thin cloud layer deep in the 
photosphere.
 \label{fig4}}

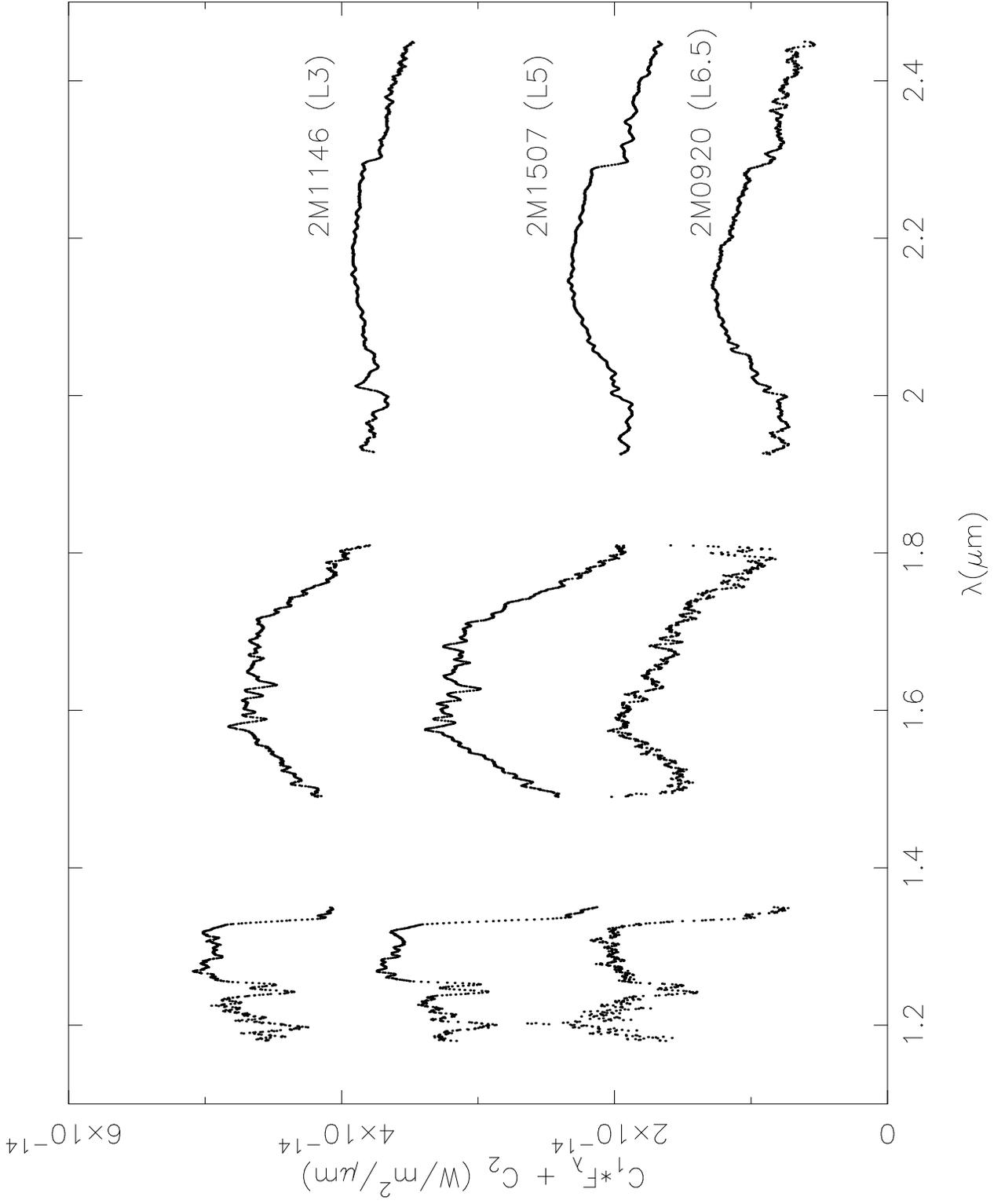
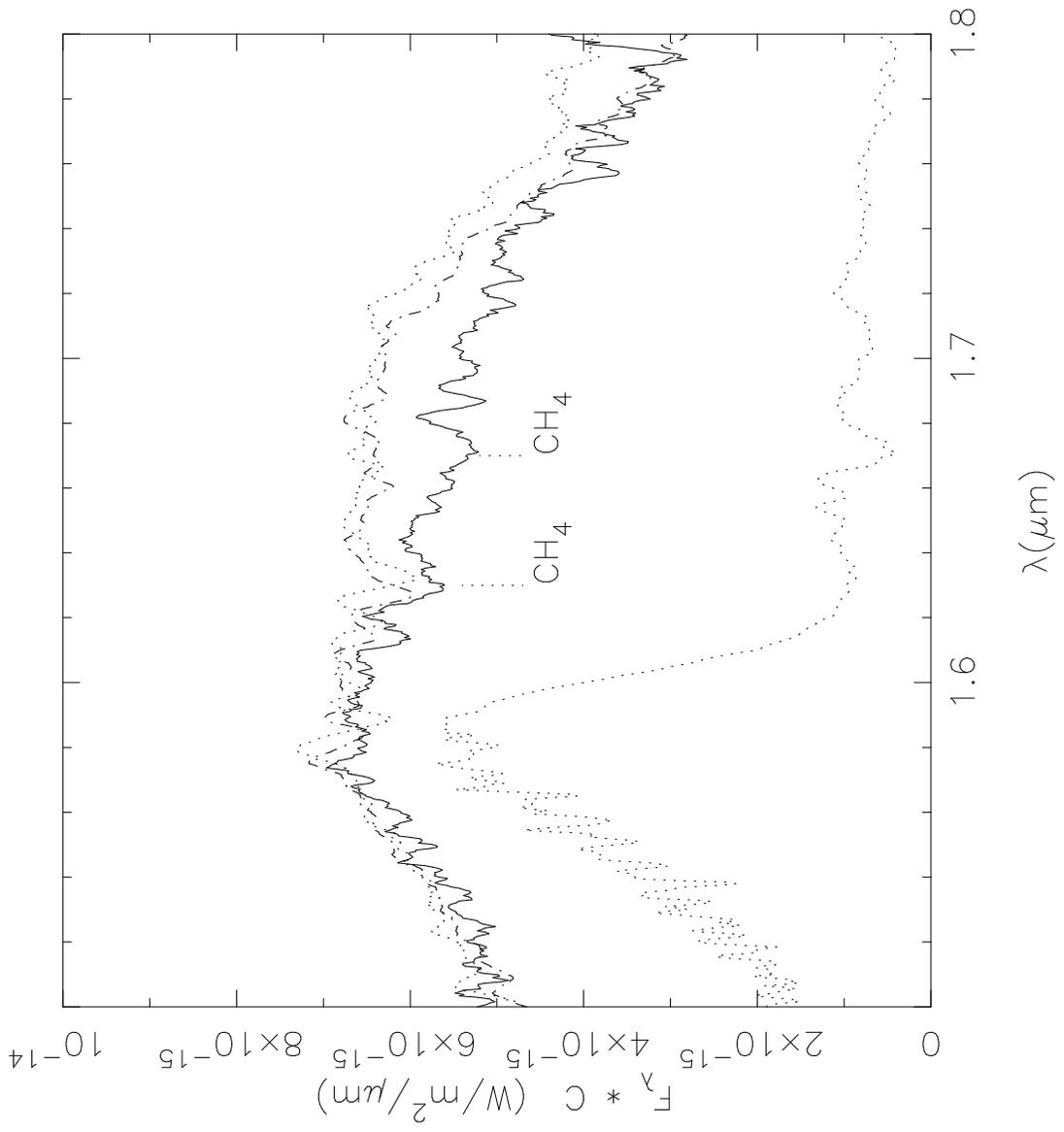
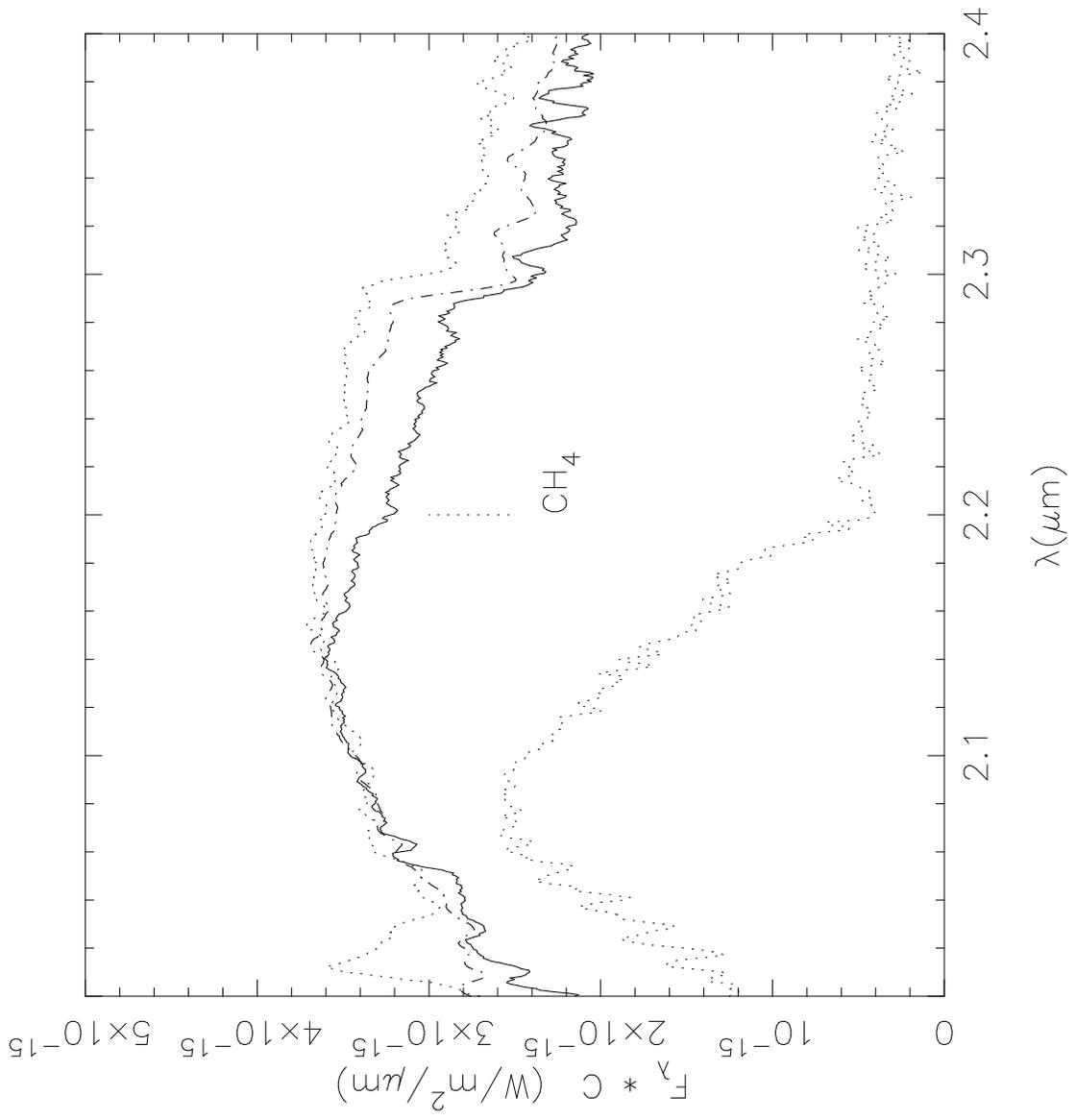
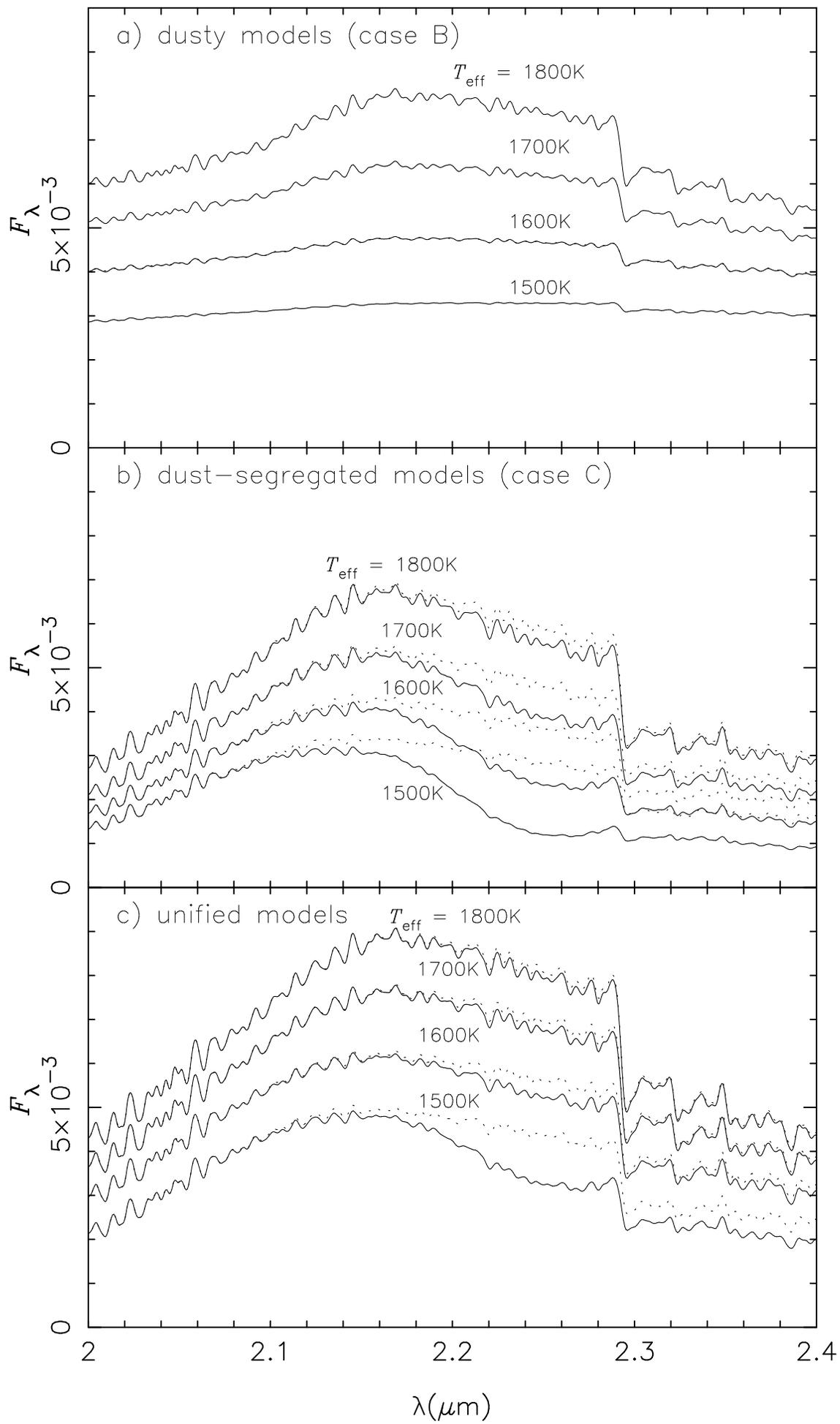
\end{document}